\documentclass[sigconf,authorversion,nonacm]{acmart}

\usepackage[utf8]{inputenc}
\usepackage{hyperref}
\usepackage{xcolor}
\usepackage{graphicx}
\usepackage{listings}
\usepackage{url}
\usepackage{soul}
\definecolor{codegreen}{rgb}{0,0.6,0}
\definecolor{codegray}{rgb}{0.5,0.5,0.5}
\definecolor{codepurple}{rgb}{0.26,0.46,0.63}
\definecolor{backcolour}{rgb}{0.95,0.95,0.95}
\usepackage{siunitx}

\lstdefinestyle{mystyle}{
    backgroundcolor=\color{backcolour},   
    commentstyle=\color{codegreen},
    keywordstyle=\color{codepurple},
    numberstyle=\tiny\color{codegray},
    stringstyle=\color{codepurple},
    basicstyle=\ttfamily\footnotesize,
    breakatwhitespace=false,         
    breaklines=true,                 
    captionpos=b,                    
    keepspaces=true,                 
    numbers=left,                    
    numbersep=5pt,                  
    showspaces=false,                
    showstringspaces=false,
    showtabs=false,                  
    tabsize=2
}

\title{Fingerprinting and Building Large Reproducible Datasets}

\author{Romain Lefeuvre}
\affiliation{
 \institution{University of Rennes}
 \country{France}}
 \email{romain.lefeuvre@inria.fr}

 \author{Jessie Galasso}
\affiliation{
 \institution{DIRO, Université de Montréal}
 \country{Canada}}
 \email{jessie.galasso-carbonnel@umontreal.ca}

 \author{Benoit Combemale}
\affiliation{
 \institution{University of Rennes}
 \country{France}}
 \email{benoit.combemale@irisa.fr}

\author{Houari Sahraoui}
\affiliation{
 \institution{DIRO, Université de Montréal}
 \country{Canada}}
 \email{sahraouh@iro.umontreal.ca}
 
 \author{Stefano Zacchiroli}
\affiliation{
 \institution{LTCI, Télécom Paris, Institut Polytechnique de Paris}
 \country{France}}
 \email{stefano.zacchiroli@telecom-paris.fr}

\begin{document}

\begin{abstract}

Obtaining a relevant dataset is central to conducting empirical studies in software engineering. 
However, in the context of mining software repositories, the lack of appropriate tooling for large scale mining tasks hinders the creation of new datasets. 
Moreover, limitations related to data sources that change over time (e.g., code bases) and the lack of documentation of extraction processes make it difficult to reproduce datasets over time.
This threatens the quality and reproducibility of empirical studies.  

In this paper, we propose a tool-supported approach facilitating the creation of large tailored datasets while ensuring their reproducibility.  
We leveraged all the sources feeding the Software Heritage append-only archive which are accessible through a unified programming interface to outline a reproducible and generic extraction process.
We propose a way to define a unique fingerprint to characterize a dataset which, when provided to the extraction process, ensures that the same dataset will be extracted. 

We demonstrate the feasibility of our approach by implementing a prototype.
We show how it can help reduce the limitations researchers face when creating or reproducing datasets.
\end{abstract}


\maketitle

\section{Introduction}
\label{introduction}

Empirical research in software engineering has experienced significant growth over the past two decades~\cite{zhang2018empirical}. In addition to the important impact of dedicated scientific venues such as MSR\footnote{https://www.msrconf.org/} and EMSE\footnote{https://www.springer.com/journal/10664/}, the proportion of papers applying empirical techniques has increased significantly in all  major software engineering venues.
Moreover, all the major conferences and journals in the field now consider reproducibility\footnote{According to the terminology used by ACM, we use in this paper the term \emph{reproducibility} to refer to the fact that the measurement can be obtained with stated precision by a different team, a different measuring system, in a different location on multiple trials~\cite{barba2018terminologies}.} to be a major evaluation factor of the submitted research results with rigorous replication guidelines~\cite{shull2008role, juristo2012replication, da2014replication}.
At the same time, much effort has been put into providing benchmarks to facilitate the evaluation of research contributions and their comparison to the current state of the art. The corresponding datasets cover several application domains such as Android apps~\cite{allix2016androzoo} and/or target specific problems such as code review~\cite{wang2021can}. In general, those datasets contain code elements and other data derived from the code that characterizes the internal properties of those elements in the form of metrics or abstract representations. They can also contain data that characterizes external properties of the code elements like, e.g., bug reports.

Generally speaking, empirical studies in software engineering follow three common steps: select relevant repositories, extract the necessary data from these repositories, and finally analyze this data to answer the research questions~\cite{vidoni2022systematic}. 
While the extracted data (\textit{refined} dataset) is strongly tied to the conducted study, the selection of repositories (\textit{raw} dataset) may be more prone to be reused as the first step of replications or other studies.
That is, different studies may extract their refined datasets from the same raw dataset.

In the context of code repositories, building reproducible raw datasets is difficult for two main reasons. First, extracting large-scale datasets for specific purposes from code forges is resource-intensive, and in most of the cases, a laborious endeavor. 
Second and more importantly, the content of repositories changes over time, up to several times a day. This makes it difficult to reproduce the same dataset over time, even when using the same extraction process.

In this paper, we propose to characterize a raw dataset in a unique way, through a fingerprint composed of a query and a timestamp.
While the query defines what constraints a repository must verify to be part of the dataset, the timestamp sets the state of the data sources which are mined to build the dataset. 
In addition, we define an extraction process which enables to retrieve from such fingerprint the same dataset at any point in time, hence ensuring its reproducibility.
We propose a generic approach that can be applied to any software forge or meta-forge as long as it guarantees immutability. 
We implement our approach on top of the Software Heritage archive which is to the best of our knowledge the sole meta-forge providing such property. Software Heritage~\cite{di2017software} stores hundreds of millions open source projects with their development histories and make them available through a unified programming interface.
Software Heritage, that we will not present in detail, is therefore used as an existing and independent platform to implement our prototype.

We show that using fingerprints coupled with our extraction process overcomes limitations faced by researchers when building, reusing or reproducing a dataset composed of software repositories.
Our approach enables any researcher to compare their work to different approaches on exactly the same data without having to reimplement those approaches or executing them on the datasets.

We illustrate our approach through a case study about open source Android applications mined from Software Heritage.
We developed a prototype and we demonstrate its ability to build a large dataset from various origins, still with a unified interface.
We show that variations in time in the fingerprint lead to different versions of a dataset.
We also test if our implemented approach is deterministic and if it can retrieve the same dataset from a given fingerprint at different points in time.
The open source implementation of the prototype is available together with the entire replication package on Zenodo.\footnote{https://doi.org/10.5281/zenodo.7989955}

The rest of this paper is organized as follows. Section~\ref{sec:motivations} discusses the limitations of retrieving datasets of code repositories through a running example. Section~\ref{sec:soft_heritage} provides background on Software Heritage and the associated features that that we use in this work. Our fingerprinting technique is described in Section~\ref{sec:approach} together with the fingerprint-based extraction process in Section~\ref{sec:operationalization}. We illustrate our approach through a case study in Section~\ref{sec:case_study}. Before concluding, we discuss related work in Section~\ref{sec:related_work}.

 \section{Motivations}
\label{sec:motivations}

In this section, we stress several limitations researchers may face during the process of acquiring a raw dataset composed of software repositories. 
We do not focus on the techniques to extract data from these repositories, but on how to obtain a curated list of relevant software repositories in the first place. 
We rely on a fictional illustrative example of a study targeting the development of modern and active open source Android applications.
To do so, one may be interested in analyzing the code in open source repositories of \emph{Android applications} which have a \emph{creation date not prior to 2015} and \emph{at least 1000 commits}.
These three selection criteria are convenient for identifying limitations, because they help cover a wide spectrum of potential obstacles by requiring to examine repository data (detecting Android applications through the presence of certain files in the repository, namely \texttt{AndroidManifest.xml}), repository metadata (creation date), and perform join operations (number of commits in the history).
In what follows, we discuss limitations which can arise in three scenarios: reusing an existing dataset, reproducing an existing dataset, and creating a new dataset.

\subsection{Limitations when Reusing Existing Datasets}

We found two refined datasets in the literature which could be useful for our case: AndroZooOpen~\cite{liu2020androzooopen} and AndroidTimeMachine~\cite{geiger2018graph}. 
AndroZooOpen provides metadata of open source Android applications, while AndroidTimeMachine gathers Android applications' commit history. 
If one wants to perform another kind of analysis on Android applications (e.g., source code analysis), these two refined datasets are not adapted. Nevertheless, their raw datasets could be reused and built upon.

AndroZooOpen~\cite{liu2020androzooopen}, published in 2020, refers to \num{46523} repositories of Android applications gathered from GitHub and F-Droid.
The dataset takes the form of a collection of CSV files documenting different types of metadata retrieved from GitHub and Google Play, as well as other artifacts (e.g., the APK retrieved from AndroZoo~\cite{allix2016androzoo}).
F-Droid\footnote{\url{https://f-droid.org/}, accessed 2023-01-18} is an app store devoted to distribute open source Android applications and information about them, including URLs towards upstream repositories containing their source code. 
All listed applications were thus considered relevant to be included in this dataset.
Identifying repositories containing the code of Android applications on GitHub is less straightforward. 
First, the authors searched on GitHub all repositories categorized under the \emph{Android} topic.
Then, they cloned these repositories and analyzed their files: they retained only the repositories containing both a main launcher \texttt{Activity.java} file and a file \texttt{AndroidManifest.xml}, which characterize Android applications.

AndroidTimeMachine\footnote{\url{https://androidtimemachine.github.io/}}~\cite{geiger2018graph} is a dataset of commit history of real-world Android apps taken from GitHub.
It combines the GitHub information for \num{8432} repositories with metadata from the Google Play Store. 

As for the previous dataset, the authors had to identify which repositories on GitHub were presenting source code of Android applications.
However, they use a different strategy.
Rather than relying on GitHub topics to filter repositories, they directly identified all repositories containing a file \texttt{AndroidManifest.xml}.

Also, instead of using the GitHub API, they exploited a GitHub mirror available on BigQuery\footnote{\url{https://console.cloud.google.com/marketplace/product/github/github-repos}} to perform their search.

\bigskip

\textbf{Limitation RU-1: links point towards resources which can be altered.}
Both existing raw datasets contain the URLs where the repositories were accessed.
However, we noticed that some of them point to projects which were deleted since the authors performed their selection.
Also, even if the projects still exist, their history may have been modified (e.g., by using \texttt{git rebase}, \texttt{git push --force} or equivalent): if so, it is impossible to retrieve the state of the repository at the time of initial dataset selection.
For instance, version control system metadata or its commit history may be different.
Therefore, \emph{it is not possible to ensure the reproducibility of a raw dataset by providing the URLs of the selected repositories, because they do not guarantee that they will still be accessible and their history preserved in the future}. Providing links towards a mirror (such as the one used by AndroidTimeMachine) may mitigate this issue. However, lack of information regarding how public mirrors evolve over time can also be a limitation (see RP-3).

Several works highlighted the necessity of providing timestamps to ensure dataset reproducibility~\cite{tutko2022software, vial2019reflections, kalliamvakou2016depth}, because they help identify which state of the repository was retrieved at the time of the selection. Timestamps are however \emph{not sufficient} when repositories are not accessible anymore, or if their history have been modified, because they are not \emph{persistent intrinsic identifiers}~\cite{DBLP:journals/cse/CosmoGZ20} based on the content of the referenced artifacts (which would correspond to the entire version control repositories, in our example).

This limitation can be generalized to every dataset which include links towards resources which can change over time. For instance, AndroidTimeMachine provides links towards the Google Play pages of some of its repositories, but in December 2022, only 30\% of these links were not producing a 404 error.

It is noteworthy that the authors of the AndroidTimeMachine dataset provide snapshots of the repositories at the time of the dataset creation, which mitigates the previous limitation. However, this may not be possible for all datasets: this solution requires a consequent storage capacity and sharing facility for large datasets, which are more and more prevalent in the recent literature due to the ever-growing popularity of machine learning approaches for software sciences. 
\vspace{-1 em}
\subsection{Limitations when Reproducing an Existing Dataset}

Even when data sources do not change over time, reproducing the steps for selecting the repositories may be necessary. 
This task is especially important in the context of reproducing empirical studies and for benchmarking.
We faced two limitations when attempting to reproduce the selection processes described in AndroZooOpen and AndroidTimeMachine.
  \newline
  
\textbf{Limitation RP-2: the selection process is not systematic and/or not clearly defined.}
To retrieve all repositories matching the \textit{Android} topic, the authors of the first dataset defined what they called a divide-and-conquer search strategy to bypass the limitations of the GitHub's search functionality.
Indeed, even if the number of repositories matched by a search query is indicated, only the \num{1000} first results are actually returned by the API.
Consequently, they divided the initial query into several more specific queries (e.g., ``all repositories with one star created before 2018 categorized in the \textit{Android} topic'') to reduce the number of matched repositories.
If one of this query matched a number of repositories superior to \num{1000}, they split the query again.
This strategy necessitates manual efforts to inspect the results of queries and to refine them, while ensuring that the set of queries is complete (i.e., they cover all the repositories).
Because the number of matched repositories for each query will evolve over time, in case the process needs to be redone, some queries used in this study will have to be manually refined again.
This process, while overcoming the limitations imposed by GitHub's API, is time-consuming, error-prone, and require manual efforts and validation from experts to define adapted queries.
The authors of AndroidTimeMachine provide the query and a link toward the BigQuery dataset on which they apply their query.
Such a formal way to express the selection and a systematic way to apply it should be considered to limit ambiguity and mistakes when reproducing a selection process.
A recent study found out that only 17\% of MSR papers describe a systematic selection process~\cite{vidoni2022systematic}.

\bigskip

\textbf{Limitation RP-3: data sources are not reliable.}
The authors of the second dataset, whose selection process relies on BigQuery, provides the query they used and how to re-run it.
To the best of our knowledge, few information are available concerning the GitHub mirror hosted on Google BigQuery.
The 3M snapshots in the mirror are GitHub repositories associated with an open source license, but we found little information regarding how representative they are of GitHub.
Also, the mirror is updated weekly, but no further information is provided on how this update affects the dataset, e.g., if it is append-only.
To date, we are not able to ensure that a query, even considering a timestamp, will yield to the same result over time, and thus if repository selection processes relying on this data source can build reproducible datasets and under which conditions.

\subsection{Limitations when Creating new Datasets}

In the case where one cannot rely on existing datasets, they may build a new one tailored to their needs.

\bigskip

\textbf{Limitation C-4: forges are heterogeneous.} 

Because of the heterogeneity of the available data sources, the definition and implementation of selection processes have to be adapted to consider the differences in available APIs, metadata. and limitations of each platform. This makes it difficult to include diverse data sources in a dataset, and consequently, researchers tend to rely upon the forge that contains the most source code and offers the best/easiest to use API for a given task.
For instance, both existing datasets use GitHub as their main data source. AndroZooOpen also considers a small existing dataset, F-Droid, although these repositories consist of roughly 4\% of their final dataset. GitHub is the platform with the most repositories and users~\cite{tutko2022software}, and  numerous tools are available to help practitioners mine GitHub data~\cite{dyer2015boa, gousios2013ghtorent}, which makes it the main data source for 67\% of MSR papers~\cite{vidoni2022systematic}. Although focusing on the prevalent forge is understandable, it induces a bias which might exclude a significant part of the objects of study. Code forges do not have the same features and are used differently by varied communities. For instance, GitLab offers different continuous integration and delivery features than GitHub, and may attract different kind of software projects than GitHub or SourceForge.

\bigskip

\textbf{Limitation C-5: forges do not provide appropriate tooling for large scale mining.}
Forges usually expose APIs mostly designed to meet the needs of the industry, allowing DevOps engineers to access repository data for automation purpose (e.g. continuous integration, dashboards).
These APIs may enable the access to the data model of a specific project, or provide search functionalities.
However, these features have many limitations which makes their usage for large-scale repository mining challenging. 

If we were to use GitHub to select repositories for our running example, we would need to define two subqueries.
The first one---\emph{``repositories of android applications'' (SQ1)}---can be fulfilled using the code search API to find all the files named \texttt{AndroidManifest.xml} and the corresponding repositories.
The second one---\emph{``repositories which have a creation date not prior to 2015 and at least 1000 commits'' (SQ2) }---can be handled by using the GraphQL API to get the creation date and the total number of commits of each repository identified in SQ1.
GitHub is known to impose a fixed rate limit for each user.
For SQ1, between \num{3} and \num{9} million results are expected. Knowing that at most \num{100} results are returned per request, it would require to run between 30K and 90K queries. With the rate limit of 30 queries per minute for the search endpoint, it would take between 17h and 50h to complete an execution.
We estimated that SQ2 would then requires more than \num{32} days to be executed on the results of SQ1.
In addition to rate limitations, one can face operational limitations. The search endpoint returns only the first \num{1000} elements for each query, while our query matches millions of elements.\footnote{\url{https://docs.github.com/en/rest/search?apiVersion=2022-11-28}}
A common tweak is to divide massive queries using available attribute filters, such as the divide-and-conquer strategy adopted by AndroZooOpen.
This requires complex heuristics which makes their implementation  constraining. 
Finally, it appears that the total count of returned elements may differ when running the same query several times, leading to non-reproducible results. 

GitLab offers a legacy REST API as well as a GraphQL API.
However, the GitLab advanced search API 
is not available on the whole forge, and thus query such as SQ1 are not supported,\footnote{\url{https://gitlab.com/gitlab-org/gitlab/-/issues/197231}, accessed 2023-01-18} making the repository selection of our example impracticable on this forge.
 \section{Software Heritage: a Meta-Forge Supporting Large \& Reproducible Mining}
\label{sec:soft_heritage}
In collaboration with the UNESCO and initiated by the National Institute for Research in Digital Science and Technology (Inria - France), the Software Heritage project\footnote{\url{https://www.softwareheritage.org/}} is built upon the idea that source code contains a form of human knowledge and is thus a part of our heritage which is worth preserving~\cite{di2017software}.
Software Heritage (SWH) collects and preserves open source software with the aim of building a universal archive of source code along with its development history, as captured by modern version control systems. Open source software are collected regularly by crawling the main forges like Bitbucket, GitHub or GitLab. Software Heritage also allows smaller forges to be archived, for instance small GitLab instances hosted by an organisation.
Software Heritage also aims to archive research software that are omnipresent in all fields and contain scientific knowledge that must be preserved. Archiving such software is crucial for reproducibility and the accessibility of the research.
Currently (January 2023), the SWH archive\footnote{\url{https://archive.softwareheritage.org}} contains more than 186 million freely accessible projects.
In the following, we focus on the properties of Software Heritage that can be leveraged to circumvent the limitations presented in Section~\ref{sec:motivations}. For a more general introduction to SWH we refer the reader to~\cite{di2017software}.

In addition to its archiving mission, SWH aims to facilitate large scale software mining by providing relevant tools and representations of the archived data. 
The objective is to tend to an ``\emph{universal software mining, i.e., making it feasible for researchers to study the entire corpus of software commons}''. 
The use of such a meta-forge facilitates the repository mining of all the crawled forges  through a unique data model and API.
This can help to overcome limitations related to the heterogeneity of existing forges (C-4), and thus help prevent bias induced by focusing on repositories from a single data source.
However, it comes with challenges, such as the necessity for the meta-forge API to offer at least the same query features as those offered by the most used forges.  
For instance, the SWH API does not allow for the moment to perform query on the content of archived files as GitHub or GitLab do. 

The SWH archive metadata relative to the source code and the version control system (VCS) are represented in a generic way in the \emph{SWH Graph Dataset}~\cite{pietri2019software}, which is a fully-deduplicated Merkle DAG.
Thus, it enables to query in an uniform manner software artifacts coming from different data sources, possibly via different VCS (e.g., Git, SVN, Mercurial).
It also facilitates the study of legacy software which have migrated through different VCS (e.g., from CVS, to Subversion, to Git) and/or relocated to different hosting platforms (e.g., from GitHub, to gitlab.com, to a self-hosted GitLab instance).
The SWH Graph Dataset offers the concept of \emph{snapshot}, allowing to capture the \emph{mutability} of the targeted VCS.  
We believe that capturing mutability is a requirement for reproducibility: traditional VCS and forges do not ensure such property (e.g., it is possible to alter git history) which can lead to different results for the same query over time. 
SWH handles this issue by offering an append-only data model, immutable by design, where graph elements have unique, persistent identifiers and cannot be altered. 
These unique identifiers (called SWHIDs~\cite{DBLP:journals/cse/CosmoGZ20}), ensure that one can refer to resources which will not be altered silently (RU-1).
An append-only model offers some guarantees regarding the reliability of the data sources over time (RP-3). SWH is also based on a distributed and open architecture, with independent archive copies. By design, there is no single point of failure and data persistence is ensured.

Two different representations of the SWH Graph Dataset are available depending on the nature of the exploration required by the user: a columnar database and a compressed in-memory graph.
The columnar dataset~\cite{pietri2019software} is composed of a set of relational tables in Apache ORC format. 
For ease of use, they are available as a public dataset on Online Analytical Processing cloud platform such as AWS Athena or Azure DataBricks allowing to process the graph without dealing with infrastructure issues.  
The columnar version contains the most metadata and enables precise search in a straightforward way as it supports SQL-based queries.
However, it comes with limitations with regard to the graph nature of the archive. 
Indeed, large graph traversals can be challenging on columnar databases contrary to graph databases, which are designed to handle such graph data models.

A compressed version of the graph~\cite{DBLP:conf/wcre/BoldiPVZ20} is also available to facilitate in-memory treatments thanks to graph compression techniques commonly used in the field of large-graph analysis. 
This compressed graph can be used through the SWH-graph API. 
The compressed graph enables deep analysis on graphs which can became too costly on a columnar representation, such as transitive closure of a given node.
However, the compressed graph version of the dataset does not contain as much metadata as the columnar version.
Furthermore, if it is not trivial to load the entire compressed dataset in memory. While without metadata 200 GiB of RAM are enough to load the compressed graph in memory, more than 4 TiB are needed to also load all associated metadata. Hence queries that require frequent metadata access (e.g., to filenames) may be less efficient than in the columnar dataset. 
The two dataset versions are thus complementary.
It is noteworthy that, contrary to existing forges which only expose their metadata through APIs, the two versions of the SWH datasets are available as open data and can be retrieved locally by the users to be accessed directly without incurring API rate limiting.
This enables to perform complex search and filtering operations on the metadata, which would have required several consecutive queries with an API, or strategies such as the divide-and-conquer one to bypass platform limitations.
To the best of our knowledge, this dataset combination constitutes the most advanced tooling for large scale repository mining (C-5).

Having a unified representation of repositories from different data sources, as well as a single architecture designed to access and analyze them can decrease the necessary efforts for defining the selection process of repositories systematically (RP-2), by sharing, for instance, the query or program which was run against a specified unalterable version of the archive.
 \section{Approach Overview}
\label{sec:approach}

\begin{figure}
    \center \includegraphics[width=0.4\textwidth]{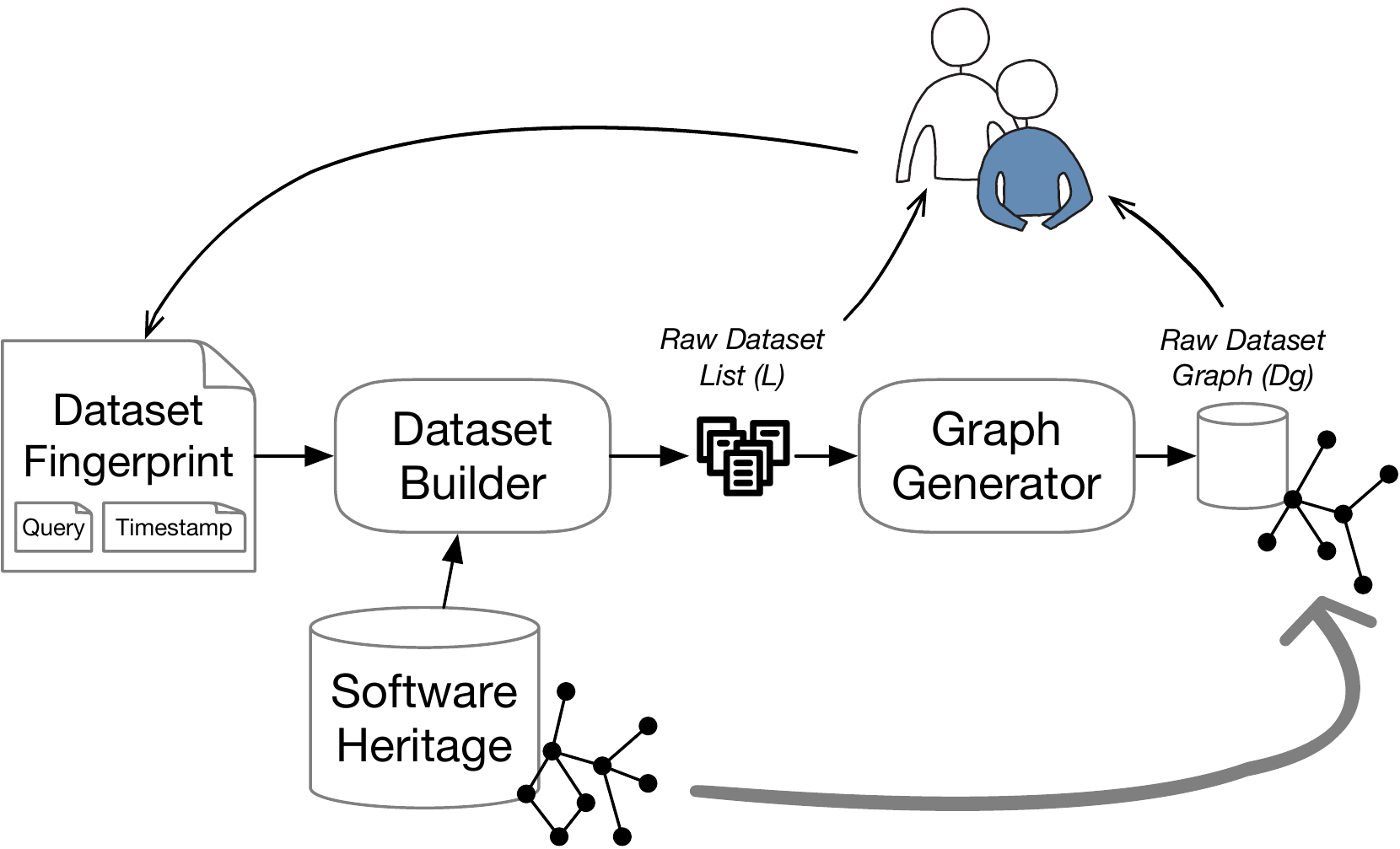}
    \vspace{-5 pt}

    \caption{Approach Overview}
    \label{fig:appraoch} 
    \vspace{-1.5 em}

\end{figure}

We propose a tool-supported approach to create and manipulate, over time, large reproducible datasets. 
The approach combines a \textit{generic selection process} (DatasetBuilder) and a \textit{dataset fingerprint }to build a dataset (cf. Figure \ref{fig:appraoch}).
We call \textit{dataset fingerprint} the minimal information characterizing a dataset such that it can be reproduced identically.
The generic selection process is built upon the Software Heritage Archive and its infrastructure, and can produce various datasets simply by providing it with different fingerprints.
By leveraging the immutability property of the SWH Graph Dataset, the proposed selection process can reproduce a dataset from a fingerprint $FP=(q,t)$ composed of a \emph{query specification} $q$ and a \emph{timestamp} $t$.

\begin{figure*}
    \center \includegraphics[width=0.8\textwidth]{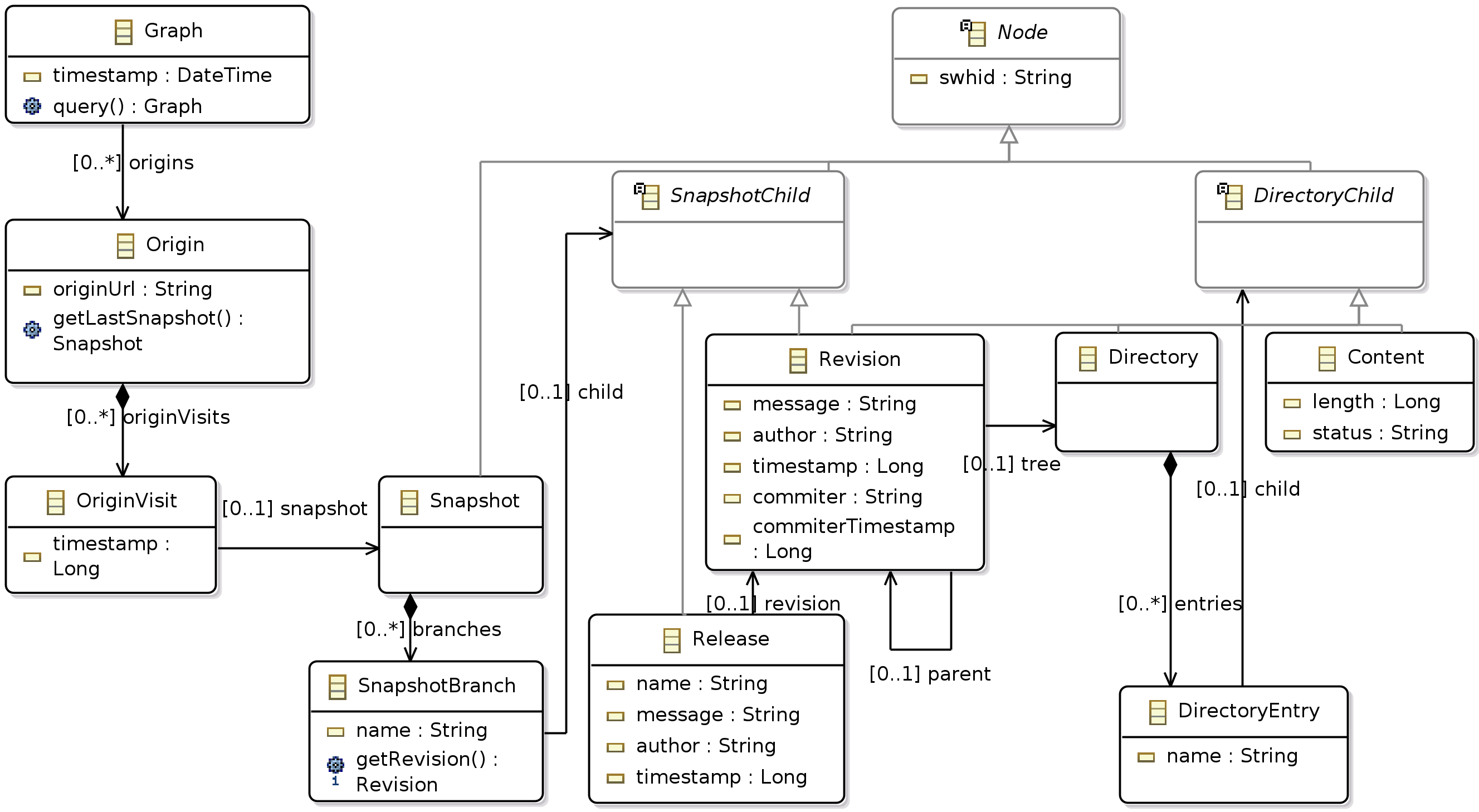}

    \caption{Object Model of the SWH Graph Dataset}
    \vspace{-5 pt}

    \label{fig:swhModel} 

\end{figure*}

The \textbf{query specification} acts as a filter by defining constraints a repository must verify to be included in the output dataset.
These constraints are expressed over the repositories' metadata as defined in the unifying domain model of the SWH Graph Dataset (presented in the form of a class diagram in Figure~\ref{fig:swhModel}, which will be detailed later).
The expressivity of possible queries is thus bounded to that model. 
For instance, it is not possible to specify a query according to the content of the artifacts, because this metadata is not included in the domain model. 
Note that the SWH model provides the commonalities among the various original forges, but misses the specificities of each one (e.g., stars in GitHub and GitLab). 
While this information is included in the related SWH archive, filtering repositories based on such specificities thus requires a post-processing operation on the extracted dataset.

The \textbf{timestamp} is a unique identifier referring to a specific version of the SWH Graph Dataset. This timestamp ensures reproducibility since each version of the SWH Graph Dataset is immutable, and the versions are append-only over the time. Hence, it is possible, at any point in time, to retrieve the dataset from the version of the SWH Graph Dataset corresponding to the timestamp, or any subsequent versions. This ensures reproducibility of the dataset even if there is further changes in the code base (e.g. Git history rewriting, branch deletion, etc.). 

The result of this process is a list of SWHIDs referring to repositories matching the query and the timestamp (cf. \emph{Raw Dataset List} in Fig. \ref{fig:appraoch}).
This list of SWHIDs can be fed to a \emph{Graph Generator} which extracts the subset of the SWH Graph Dataset corresponding to these identifiers (cf. \emph{Raw Dataset Graph} in Fig. \ref{fig:appraoch}).
Consequently, the obtained dataset of repositories can be further manipulated with the same SWH infrastructure and programming interface.
For instance, it can be used later on as input of our approach to filter out elements according to a more restrictive fingerprint. 
It can also enable to filter and download specific files, instead of cloning all the repositories to extract only a few files from them during the data extraction phase, which follows the repository selection phase~\cite{vidoni2022systematic}.

To sum up, our approach can be defined as the following function:
\[ SWHg \times FP \rightarrow L \rightarrow Dg \]
where $SWHg$ is the SWH Graph Dataset, $FP$ a dataset fingerprint, $L$ the list of origin ID's matching $FP$, and  $Dg$ the resulting dataset in the form of a subset of the SWH Graph Dataset.

Although our approach rely on SWH, we propose a generic approach that can be applied to any forges that provide an immutability property and a way to query the metadata of its repositories.
 \section{Approach Operationalization}
\label{sec:operationalization}
In this section, we propose an operationalization of our approach.
We present the implementation of a compiler which transforms a query into a Java program calling the SWH Graph API.
We first discuss our choice of the Object Constraint Language (OCL)~\cite{DBLP:conf/sfm/CabotG12} to specify the query, then how we manage timestamps.
Finally, we explain how the implemented process generates a executable program from the formers to retrieve SWHIDs of the matching repositories.

\subsection{Dataset Specification}

To formally describe a subset of the repositories included in the SWH Graph Dataset, we rely on a query expressed in a query language.
Several query languages exist which can be used for this purpose.
When selecting a query language, we considered the language expressiveness and the facility to use the language for code generation purpose.

SQL can be used with the relational version of the archive metadata, but as we said previously, this solution is not appropriate to express complex queries due to the graph nature of the archive.
Graph query languages such as \emph{GraphQL}~\cite{DBLP:conf/www/Hartig018} or \emph{Cypher}~\cite{DBLP:conf/sigmod/FrancisGGLLMPRS18} could be used on the compressed version of the graph.
However, we found that GraphQL has some limitations regarding its expressiveness (lack of conditional support, join operation or user defined function) and does not allow transitive closure.
Cypher addresses most of the limitations mentioned for GraphQL, making it a good candidate as our query description language. 
However, the lack of tooling facilitating the creation of generators or an available editor has led us to discard this language. 

OCL (Object Constraint Language) is an object query language allowing to describe constraints on an object-oriented model, which can also be used to express complex queries without side effects.
OCL is widely used in the model-driven engineering community and comes with numerous tools, such as the Eclipse OCL implementation enabling to express constraints on a UML or an Ecore model.
It also provides an editor and facilities for creating generators. 
To be able to express OCL queries on the archive metadata,   
we defined an object-oriented model of the SWH Graph Dataset (Fig.~\ref{fig:swhModel}) representing the different elements considered in the metadata schema. 
The model conforms to the structure of the compressed graph described in~\cite[Chapter 10]{pietri:tel-03515795}.
An instance of this model represents an export of the Software Heritage archive.
The \texttt{Graph} class thus represents a specific version of the Graph Dataset.
It is composed of a list of repositories and a timestamp representing the version of the export. 
The repositories are represented by the \texttt{Origin} class and are identified by their URLs.
Every time a repository is crawled by SWH, the state of the repository is captured by a timestamp and a snapshot as an \textit{OriginVisit}. 
The rest of the model is similar to the Git Merkle DAG. 
Indeed, a \texttt{Snapshot} is composed of a list of \texttt{SnapshotBranch}es pointing to a \texttt{Revision} (equivalent of a Git commit) or a \texttt{Release} (equivalent of a Git tag). 
Finally, each \texttt{Revision} is composed of a file tree and a reference to the previous \texttt{Revision}. 
The \texttt{Snapshot, Release, Revision, Directory} and \texttt{Content} classes inherit from the Node interface and are identified by SoftWare Heritage persistent IDentifiers (SWHIDs) which are guaranteed to remain resolvable over time.\footnote{\url{https://docs.softwareheritage.org/devel/swh-model/persistent-identifiers.html}} The SWHID also enables integrity check of an entire snapshot since it contains the SHA1 hash of the referenced object.

\begin{figure}[ht]
\begin{lstlisting}
import swhModel : 'platform:/resource/.../swhModel.ecore'
package swhModel
context Graph
def : query():Set(Origin) = origins->select(
	 getLastSnapshot().branches->exists(
	 	(name='refs/heads/master' or name='refs/heads/main')
		and
		/*The branch contains at least 1000 revisions */
		getRevision()->closure(parent)-> size() >1000
		and
		/*The root revision have been created since 2015 */
		getRevision().getRootRevision().commiterTimestamp>1420066800
		and
		/*The branch contains a file 'AndroidManifest.xml'*/
		getRevision().tree.entries->closure(entry:DirectoryEntry |
				if entry.child.oclIsKindOf(Directory) then
					entry.child.oclAsType(Directory)
                        .entries.oclAsSet()
				else 
					entry.oclAsSet()
				endif	
		)->exists(e:DirectoryEntry | e.name='AndroidManifest.xml')))
context Revision
 def : getRootRevision() : Revision =
	  if parent = null then self
    else parent.getRootRevision() endif
endpackage

\end{lstlisting}
\caption{The running query expressed in an OCL expression}
\label{fig:OCL_query}
\end{figure}
 
OCL allows to define methods without side effects over the classes of a given model, hence enabling to express queries on this model.
Figure~\ref{fig:OCL_query} presents an OCL query corresponding to our illustrative example about Android applications (Section~\ref{sec:motivations}).
The first line of the OCL file indicates the model: in our case, \texttt{swhModel.ecore} is the model presented in Fig.~\ref{fig:swhModel}.
We define a method named \textit{query} (line 4) which returns a set of \texttt{Origins} (repositories) when applied to an instance of the class \texttt{Graph} (\texttt{context Graph}, line 3).
In other words, this method takes a version of the SWH archive and selects repositories matching the filters defined in the query (lines 4 to 24).
We can see in line 5 that we select the origins whose last snapshot contains at least a branch matching the following conditions:
  \begin{itemize}
      \item Line 6: The branch name must be ``main'' or ``master''.
      \item Line 9: The branch must contain at least 1000 revisions, we used the closure operation on the last revision of the branch to flatten the parent revision relation. 
      \item Line 12: The root revision must have been created after 2015. We define the ``\texttt{getRootRevision()}'' operation in the \texttt{Revision} context (lines 25-31) to retrieve the first revision of a branch.
      \item Lines 15-22: The revision must contain an entry named \\``\texttt{AndroidManifest.xml}''. The file tree of the revision is flattened into a set of \textit{DirectroryEntry} with a closure operation.
  \end{itemize}

\subsection{Timestamp Management}

The SWH archive is continuously evolving over time by crawling snapshots of repositories and updating the current archive state. The model of Fig.~\ref{fig:swhModel} contains two types of timestamps. The timestamp in \texttt{OriginVisit} defines the time where the snapshot of a given repository was taken.
The timestamp in \texttt{Graph} indicates a frozen version of the Graph Dataset which contains all the \textit{OriginVisit}s taken before this timestamp. 

Both timestamps allow us to fix the state of the archive on which the query will be executed and ensure that the query output will be reproducible for a given timestamp.  
Since the SWH archive is immutable, it is theoretically possible to filter a given export to retrieve the state of a previous export and execute a \textit{query} on it. 
In other words, if we run a query \texttt{q} on an export realized at a time $t$, re-running the same query \texttt{q} on an export done at a time $t+m$ while discarding all the \textit{OriginVisit}s  added after $t$ should produce the same selection.

\subsection{Repository Selection}

To automate the selection of the repositories from the SWH archive that match the filters expressed in the OCL query, we implemented a compiler translating the constraints of the query in an optimized Java program which uses the SWH-Graph API to perform filtering operations on the SWH Graph Dataset. 

We first wrapped the SWH-Graph API to conform to the object-oriented model of Fig~\ref{fig:swhModel}.
Indeed, the SWH-Graph API is not fully object-oriented, and relies on static methods to retrieve node properties  to improve its efficiency (i.e., avoiding object creation). 
Wrapping this API to the same OO model highly facilitates the translation of the OCL query in Java code.

The compiler relies on a generator taking as input an OCL query.
First, it uses Eclipse OCL PIVOT, which provides an Xtext grammar of OCL, to build the Abstract Syntax Tree (AST) of the query.
The obtained AST is then traversed to generate the corresponding Java code.
For this, we specify in the generator one generation method for each type of nodes of the AST, defining the Java code to be produced if this type of node is encountered during the traversal.
For instance when the \textit{exists} OCL operation is used in the query, it will trigger the generation method of the \textit{IteratorExp} node type and produce Java code using \textit{stream().anyMatch()}.
Visiting the entire AST thus produces the corresponding executable Java code calling the SWH-Graph API to select repositories.

Executing this code outputs the set of SWHIDs referring to \textit{Origin}s (repositories) matching the constraints expressed in the initial query. 
Then, this set of origin IDs is possibly used to extract the sub-graph of the SWH Graph Dataset restricted to the provided origins, in both column based or compress format. 

The source code of the compiler is available on the replication package associated to this paper.\footnote{https://doi.org/10.5281/zenodo.7989955} \section{Illustrative Case Study}
\label{sec:case_study}
In this section, we report on the application of our prototype to extract a raw dataset for the illustrative example about open source Android applications.
Our main goal is to verify whether the proposed operationalization satisfies the properties of the approach presented in the Sections~\ref{sec:soft_heritage} and~\ref{sec:approach}, to overcome the limitations identified in Section~\ref{sec:motivations}. 
While the results are not generalizable as with a full evaluation, they provide valuable insights on the benefits, as well as the limitations to be addressed in the future.
We structure our observations through three research questions: 
\begin{itemize}
    \item \textbf{RQ1}: \textit{What is the impact of the temporal dimension of the fingerprint on the extracted dataset?} With this question, we want to obtain an order of magnitude of the difference and size and diversity of the extracted dataset when applying the same query with different timestamps. 
   
    \item \textbf{RQ2}: \textit{Is the implemented selection process deterministic?}  We want to verify that running the fingerprint several times results in the same dataset.

    \item \textbf{RQ3}: \textit{Is it possible to retrieve the same dataset when applying the fingerprint on different versions of the SWH archive?} With this question, we want to assess if our selection process is able to extract the same dataset overtime.
\end{itemize}

\subsection{Experimental Settings}

To answer these questions, we analyzed the results of different fingerprints ran over several export versions of the SWH Graph Dataset. 
In our case, the fingerprints have the same query (Fig.~\ref{fig:OCL_query}) and differ in their timestamps.
We consider two datasets to be equivalent if they contain the same list of repositories.

For our experiment we leverage on the 2018-09-25, 2021-03-23, 2022-04-25 and the 2022-12-07 export versions of the SWH archive. 
As SWH provides an export date with an uncertainty of a day, we define the graph exports $G_1\langle t_1\rangle, G_2\langle t_2\rangle, G_3\langle t_3\rangle$ and $G_4\langle t_4\rangle$ with the following timestamps: $t_1=$ \texttt{2018-09-24 UTC+1}, $t_2=$ \texttt{2021-03-22 UTC+1}, $t_3=$ \texttt{2022-04-24 UTC+1}, $t_4=$ \texttt{2022-12-06 UTC+1}.

To obtain an order of magnitude of the impact of the time dimension (\textbf{RQ1}) of the fingerprint, we run 3 different fingerprints sharing the same query but having a different timestamp. 
 Given $q$ our running query, we consider the three fingerprints:
\begin{center}
     $FP_1 = \langle t_1,q\rangle$    \;\;\;\;     $FP_2 = \langle t_2,q\rangle$   \;\;\;\;    $FP_3 = \langle t_3,q\rangle$ 
\end{center}
 The first experiment therefore consists in performing the following runs and comparing the different lists of Origins they returned:
\begin{center}
     $FP_1 \times G_3$    \;\;\;\;     $FP_2 \times G_3$   \;\;\;\;    $FP_3 \times G_3$ 
\end{center}

 In order to check if our prototype returns the same result for a given fingerprint, we run the same fingerprint twice on the same export version (\textbf{RQ2}). For this second experiment, we perform the following run two times and compare the returned lists of Origins: 
\begin{center}
     $FP_2 \times G_3$  
\end{center}

Finally, we checked if we obtain the same dataset with a same fingerprint overtime (\textbf{RQ3}). For this experiement, we run a given fingerprint on two export versions, and compare the two returned list of Origins:
  \begin{center}
     $FP_3 \times G_3$    \;\;\;\;     $FP_3 \times G_4$  
\end{center}

Our experiments have been realized on two different machines. Setting 1: ProLiant DL380 Gen10 Plus - Debian 11
\begin{itemize}
   \item CPU : 2  X Intel(R) Xeon(R) Gold 6342, 2.80GHz
   \item RAM :  32 X DDR4 3200 MHz 128GiB
   \item DISK : 12 x 5.8 TB SSD 
\end{itemize}
All our experiments have been executed on this non-reserved machine (other experiments were running at the same time on the machine). 
As the machine is non-exclusive we cannot estimate an exact execution time of our query in this case. 
Nevertheless, in these conditions the execution time was in average $\approx$7 hours.

Setting 2: ProLiant DL365 Gen10 Plus - Ubuntu Server 22.04.3
\begin{itemize}
    \item CPU : 2 x AMD EPYC 7543 32 core, 64 thread, 2.8GHz
    \item RAM : 512 GB
    \item DISK : 3 X 2To SSD Raid 5, 3To HDD
\end{itemize}
This second setting allowed us to perform experiments on exclusive resource and measure an execution time with all the machine resources. 
The run of $FP_3 \times G_3$ took 23 hours and 25 minutes. 
A thorough evaluation is needed to assess the computation time required for the execution. 

 \subsection{Experiments Results}

\subsubsection{\textbf{RQ1 — Impact of the temporal dimension}}
\vspace{-5 pt}
\begin{table}[ht]
\centering
\small
\caption{Variation of the temporal dimension between three fingerprints having the same query : Results of  $FP_1 \times G_3$, $FP_2 \times G_3$ and $FP_3 \times G_3$}
\vspace{-5 pt}

\begin{tabular}{ |c|c|c|c| }
\hline
 \textbf{Forge} &\textbf{FP1(2018)} & \textbf{FP2(2021)} & \textbf{FP3(2022)} \\
\hline
            github.com &    830 &  135820 & 172012 \\
            gitlab.com &      3 &      67 &   1154 \\
         bitbucket.org &      - &      76 &    106 \\
          codeberg.org &      - &      55 &     84 \\
          framagit.org &      - &      21 &     23 \\
     git.launchpad.net &      - &      10 &     14 \\
gitlab.freedesktop.org &      - &         - &     14 \\
            0xacab.org &      - &         - &      3 \\
       ...             &      - &         5 &      12 \\
       \hline
\textbf{Total} & \textbf{833} & \textbf{136054} & \textbf{173422} \\
 \hline
\end{tabular}
\label{table:G2xFP2_G2xFP1}
\end{table}

Table~\ref{table:G2xFP2_G2xFP1} summarizes the results of the execution of $FP_1$, $FP_2$ and $FP_3$ over $G_3$. 
For each run, more than 99\% of the repository Origins come from GitHub.
It is consistent with the proportion of the entire SWH Graph Dataset where more than 93\%  (cf. Table~\ref{table:G2}) of the repositories are extracted from GitHub (in April 2022). 
There is a substantial variation in the numbers of results between $FP_1$ and $FP_3$: we note an increase of 20719.0\% in only 4 years. 
Similarly, although the two last fingerprints are only 13 months apart, there is a non-negligible variation (+27,4\%). Furthermore, the evolution of the numbers of results is not uniform: the results on GitHub have increased by 26.7\% compared to 1622.4\% on GitLab.com.
These variations strengthen the importance of freezing the temporal dimension of the forges or meta-forges that we use to build a dataset. 
 
We also observe an increase of the numbers of different forges that produce results for our request, due to the continuous addition of new forges to the SWH archive. For instance, Bitbucket.org wasn't crawled in 2018 (see Table~\ref{table:G2}) while it produces results for $FP_2$ and $FP_3$. Thus, it is simply a matter of changing the timestamp of the fingerprint to update a dataset,  avoiding the need to manage the various APIs of the new forges.
\begin{table}[ht]
\small
\centering
\caption{Evolution of the number of forges and the number of repositories per forges  crawled by SWH : Total numbers of Origins per forge in $G_1$, $G_3$ and $G_4$}
\vspace{-5 pt}

\begin{tabular}{ |c|c|c|c| }
\hline
 \textbf{Forge} & \textbf{G1(2018-09)} & \textbf{G3(2022-04)} & \textbf{G4(2022-12)} \\
\hline
                github.com &  56404072 &  164713349 & 177810125 \\
        gitlab.com &    537541 &    4279918 &   4786089 \\
     bitbucket.org &           - &    2566198 &   2589887 \\
     www.npmjs.com &           - &    1835697 &   1835697 \\
          pypi.org &     63860 &     467142 &    530254 \\
code.launchpad.net &           - &          1 &    334081 \\
   git.code.sf.net &         1 &     183172 &    183200 \\
     gitorious.org &    116360 &     120380 &    120380 \\
   svn.code.sf.net &           - &     102765 &    102901 \\
               ... &    726860 &    1177007 &   1300455 \\

 \hline
\textbf{Total} & \textbf{57848694} & \textbf{175445629}& \textbf{189593069} \\
 \hline
\end{tabular}
\label{table:G2}
\end{table}
\vspace{-1 em}

 \subsubsection{\textbf{RQ2 — Determinism of the prototype}}

Running the fingerprint $FP_3$ on  $G_3$  twice resulted in identical results as those shown in Table~\ref{table:G2xFP2_G2xFP1}.
This observation suggests that the implemented selection process is indeed deterministic. 
However, a more thorough evaluation is necessary to attest that our prototype verifies this property in all cases. 

\subsubsection{\textbf{RQ3 — Reproducing a dataset overtime}}

We first ran the fingerprint $FP_3$ on the export version $G_3$ which shares its timestamp $t_3$: this emulates that $FP_3$ was ran on the latest version of the archive.
Then, we ran the same fingerprint on the export version $G_4$ which has a higher timestamp, corresponding to a version later in time.
Table~\ref{table:RQ3} shows that executing the fingerprint on a version of the archive which is superior to the fingerprint timestamp allows to reproduce the results with a precision of 96.8\%.
These results indicate that our prototype is nearly capable of entirely capturing the temporal dimension. 
Therefore, the prototype is close to being able to obtain identical results when executing on newer export versions fingerprints which were originally run on previous states of the archive.
The 3.2\% uncertainty can be explained both by the implementation of our prototype, but also by the limitations of Software Heritage which we will discuss in Section~\ref{sec:discussion}.
It is therefore preferable, in the current state of the prototype, to run a fingerprint on the export version corresponding to the same timestamp to ensure reproducibility.

\begin{table}[ht]
\centering
\caption{Execution of the same fingerprint on a different export of the SWH Graph Dataset : Result of $FP_3 \times G_3$ and $FP_3 \times G_4$}
\vspace{-5 pt}

\begin{tabular}{ |c|c|c|c| }
\hline
 \textbf{Forge} & \textbf{FP3 X G3} & \textbf{FP3 X G4} & \textbf{Difference (\%)} \\
\hline
       github.com & 172012 & 166630 &            -3.2 \\
   gitlab.com &   1154 &   1223 &             5.6 \\
bitbucket.org &    106 &    102 &            -3.9 \\
 codeberg.org &     84 &     84 &             0.0 \\
 framagit.org &     23 &     22 &            -4.5 \\
          ... &     43 &     38 &           -13.2 \\
 \hline
\textbf{Total} & \textbf{173422} & \textbf{168099}& \textbf{-3.2} \\
 \hline
\end{tabular}
\label{table:RQ3}
\vspace{-1.5 em}

\end{table}
 \subsection{Discussion} 
\label{sec:discussion}
The operationalization is based on Software Heritage tools which have their own limitations.

\paragraph{Internal limitations}
The SWH crawling process requires significant resources to create and maintain up to date an archive of publicly available software. This process rise multiple challenge such as the constraints imposed by forges API (rate limit, expressivity \& heterogeneity of the API). As a consequence, the modification performed on a repository are crawled periodically. Our operationalization is based on a fixed and reproducible state of the SWH archive. There is no guarantee that the current state of all the repositories of a forge have been crawled at a given time.

Finally, the SWH Graph Dataset is not built incrementally and needs to be built from scratch to be updated. Thus, there is no real-time version of the SWH Graph Dataset describing the current state of the SWH archive, but rather periodic exports are made available (on a yearly basis, at the time of writing). 

\paragraph{External limitations}
Software Heritage, like all content providers, is subject to regulations. Take down notices can be therefore submitted for various reasons (copyright, GDPR compliance on personal data deletion) requiring the removal of content from the archive. 
 \section{Related Work}
\label{sec:related_work}
\paragraph{Platforms and tools for mining software repositories}

GitHub proposes a REST API with a \textit{search endpoint} to search for specific items---including repositories---meeting certain criteria.
Each search may present up to \num{1000} results: the API is thus not made to retrieve \emph{all} items meeting the given criteria and may be too limited for the purpose of selecting repositories for MSR-based studies. According to Cosentino et al.~\cite{cosentino2016findings}, recurring reported limitations of GitHub API in MSR studies include limited quota and events not accurately returned.
To overcome these limitations, third party services were proposed to ease the mining of GitHub repositories through dataset mirrors.
GH Archive\footnote{\url{https://www.gharchive.org/}} records all public events from GitHub and makes them accessible for large scale analysis.
It is updated each hour and the dataset is available through downloadable archives and on BigQuery.
Similar to GH Archive, GHTorrent~\cite{gousios2013ghtorent} records public events of GitHub retrieved through the GitHub API and redistributes the gathered metadata in a SQL database.
Since 2019 GHTorrent is only sporadically maintained, with a most recent data dump dating back to March 2021.
GitHub Activity Data is a snapshot of the content of 3M repositories of GitHub available on Big Query.\footnote{\url{https://hoffa.medium.com/github-on-bigquery-analyze-all-the-code-b3576fd2b150}}

Boa~\cite{dyer2015boa} is a domain specific programming language for defining analysis tasks in the context of mining software repositories.
Boa comes with an infrastructure which compiles a Boa program to be run on distributed clusters to improve efficiency.
The defined analysis task is run on repositories whose information is locally cached.
Several datasets are available corresponding to difference language and forge ecosystems, but they are updated sporadically.
The most recent ``large'' dataset encompassing a significant GitHub subset dates back to October 2019 and covers 7.8 million public repositories.\footnote{\url{}https://boa.cs.iastate.edu/stats/}
No guarantee of long-term availability of the platform or the data hosted on it are provided.

\paragraph{Limitations and best practices to create and share raw datasets of code repositories}

Vidoni~\cite{vidoni2022systematic} presents a systematic literature review investigating MSR-based studies which enables to identify recurring limitations in current practices.
The author then proposes guidelines to improve MSR-based studies through the definition of a systematic process inspired by evidence-based software engineering.
The scope of their analysis is wider than ours, because they focus on the processes of selecting repositories, extracting data from these repositories and mining information for the study, while we focused on the first step of obtaining the raw dataset.
For this first step of MSR, we discuss more limitations and propose, instead of guidelines, a systematic process for selecting repositories in a reproducible way.

Vial et al.~\cite{vial2019reflections} investigate data quality issues in the context of digital trace data (DTDs).
They observe that DTDs are usually gathered form data sources over which researchers have little to no control, thus making their quality difficult to ascertain. This observation resonated with our limitation RP-3 about the unreliability of data sources. 
They do not propose a process to ensure DTDs’ quality, but encourage researchers to clearly describe this data through the Seven Ws (what, when, where, how, who, which and why) of data quality as defined by Marsden and Pingry~\cite{marsden2018numerical}, such that the reader can assess their quality.

ACM's empirical standards for mining repository studies\footnote{\url{https://acmsigsoft.github.io/EmpiricalStandards/docs/?standard=RepositoryMining}} include defining how and why repositories were selected, along with the detailed acquisition process, among the essential attributes which should be documented in studies.

Tutko et al.~\cite{tutko2022software} presents a systematic literature review about how software repositories are mined in MSR-based studies. Their findings include the non-reproducibility of most of the studied papers (e.g., lack details regarding the data selection and extraction processes, missing timestamps) and they propose a list of information which should be included in such studies to improve reproducibility.
Cosentino et al.~\cite{cosentino2016findings} present a systematic literature review about how research papers have addressed the task of mining repositories focused on GitHub.
They derive several concerns about data collection, size of the used datasets which are usually not large enough and replicability.
They found out that most of the papers report issues with the limitations of the GitHub API and with the data available from third party services such as BOA and GHTorrent, which align with the limitations C-5 and RP-3, respectively.

 \section{Conclusion and perspectives}
\label{sec:conclusion}
In this paper, we studied the problem of building reproducible datasets composed of software repositories.
We first identified five limitations researchers can face when obtaining such datasets, either by reusing existing ones, reproducing an existing selection process or creating a new dataset from scratch.
We considered from this angle the Software Heritage archive (SWH), assessing which of its interesting properties can be leveraged to overcome the identified limitations.
We introduced a new approach of dataset fingerprinting to characterize datasets with a pair (query, timestamp).
We implemented the proposed approach using the OCL language to specify the query, and a compiler which generates a Java program that uses the SWH API to extract all the repositories matching the provided query from the SWH archive.

Several perspectives can be envisioned concerning the operationalization of our approach.
One of them is the possibility to optimize a given OCL query. 
For instance, the predicate operands in an ``AND'' logical expression can be reordered to ensure that the least resource-intensive requests are executed first. 
Other heuristics could leverage the implementation choices made in Software Heritage, for instance, to better orchestrate the memory access according to the nature of the storage where the attributes in the query are stored. 
Indeed, labels are stored on disk while node types are always stored in RAM, making them faster to read.

The operationalization relies on the Java API of the SWH (compressed) graph dataset, which enables complex and efficient graph traversal operations. 
The graph is divided in several parts, one representing its structure, and the others the rest of the metadata.
To achieve the best performances, both the graph and associated metadata ($\approx$\,4.5 TiB) must be loaded into memory, which requires a substantial infrastructure. 
If the available infrastructure is not powerful enough, it is possible to load only the graph structure in memory, and to access the metadata from the disk.
In this case, the use of the column based version becomes more efficient for some processing requiring many disk accesses. 
Therefore, a hybrid approach based on both the compressed and columnar versions of the dataset can be envisioned to accelerate the evaluation of a given query.

Another perspective is to add a hash of the resulting dataset to the fingerprint.

Such hash can attest that two dataset versions are strictly identical, mitigating the impact of take down notices.

Exploring other technology stacks for the approach operationalization could be useful to better fits the needs and expertises of different users.
One could imagine replacing the OCL query language to describe the selected dataset by a domain specific language like Boa, which was designed for tasks related to mining software repository. Finally, running a large scale evaluation on several different fingerprints over different exports of the SWH Graph Dataset would allow us to verify and generalize our current observations.

\bibliographystyle{ACM-Reference-Format}
\bibliography{references.bib}
\end{document}